\documentclass[12pt]{article}

\usepackage{times}
\usepackage{graphicx}
\usepackage{amsmath}
\usepackage{amssymb}

\usepackage{xcolor}

\title{Experimental Data from a Quantum Computer Verifies the Generalized Pauli Exclusion Principle}

\author
{Scott E. Smart,$^{1}$ David I. Schuster,$^{2}$ and David A. Mazziotti $^{1\ast}$\\
\\
\normalsize{$^{1}$Department of Chemistry and The James Franck Institute}\\
\normalsize{The University of Chicago, Chicago, IL 60637, USA}\\
\normalsize{$^{2}$Department of Physics and The James Franck Institute}\\
\normalsize{The University of Chicago, Chicago, IL 60637, USA}\\
\\
\normalsize{$^\ast$To whom correspondence should be addressed; E-mail: damazz@uchicago.edu}}

\date{}

\begin{document}

\baselineskip24pt

\maketitle


\textbf{
``What are the consequences ... that Fermi particles cannot get into the
same state ... " R. P. Feynman wrote of the Pauli exclusion principle,
``In fact, almost all the peculiarities of the material world hinge on
this wonderful fact."   In 1972 Borland and Dennis showed that there
exist powerful constraints beyond the Pauli exclusion principle on the
orbital occupations of Fermi particles, providing important restrictions
on quantum correlation and entanglement.  Here we use computations on
quantum computers to experimentally verify the existence of these additional
constraints.  Quantum many-fermion states are randomly prepared on the
quantum computer and tested for constraint violations.  Measurements
show no violation and confirm the generalized Pauli exclusion principle
with an error of one part in one quintillion.}

\section*{Introduction}

While performing calculations with classical computers at IBM, Borland and Dennis discovered something unexpected and surprising about the electronic structure of atoms and molecules\cite{Borland1972}.  In 1926 Pauli had observed that no more than a single electron can occupy a given one-electron quantum state known as a spin orbital\cite{1925Pauli}.  Formally, the Pauli exclusion principle implies that the spin-orbital occupations are rigorously bounded by zero and one. In their calculations Borland and Dennis discovered, however, that even in three-electron atoms and molecules there are additional constraints beyond the well-known exclusion principle. In 2006 Klyachko (and in 2008 with Altunbulak) presented a systematic mathematical procedure for generating these constraints for potentially arbitrary numbers of electrons and orbitals\cite{Klyachko2006,Altunbulak2008}. These inequalities, which have become known as generalized Pauli constraints\cite{Schilling2013, Benavides-Riveros2015,Chakraborty2014,Schilling2017}, provide new insights into the electron structure of many-electron atoms and molecules\cite{Chakraborty2014, Theophilou2015,Mazziotti2016,SB2016,Erdahl2007}, the limitations of entanglement as a resource for quantum control\cite{Rabitz2004,Walter2012}, as well as the fundamental distinctions between open and closed quantum systems\cite{Chakraborty2015}.

In this work we use quantum states prepared on a quantum computer to provide experimental verification of the generalized Pauli constraints.  Quantum computers differ from classical computers in that quantum states can be prepared on the quantum computer.  Algorithms which utilize these quantum states promise a large computational advantage over known classical algorithms for solving critically important problems such as integer factorization, eigenvalues estimation, and fermionic simulation\cite{Shor1995, Abrams1997, Bravyi2000, Kais2018}. Rapid advances in quantum hardware and hybrid classical-quantum algorithms have led to multi-qubit experimental implementations\cite{Kandala2017,Naik2017,Gambetta2015}.   We first randomly prepare the quantum state of a \textcolor{black}{3-electron} system and second measure the occupations of the \textcolor{black}{natural} orbitals.  \textcolor{black}{The natural orbitals are the eigenfunctions of the 1-electron reduced density matrix (1-RDM) that is defined by integrating the many-electron density matrix over the coordinates of all electrons except one
\begin{equation}
^{1} D(1;{\bar 1}) = \int{ \Psi(123) \Psi^{*}({\bar 1}23) d2 d3}
\end{equation}}
in which $\Psi(123)$ is the 3-electron wave function.  The two-step process is repeated many times on the quantum computer to explore all possible physically realizable orbital occupations.  \textcolor{black}{Because of the generalized Pauli constraints} a large convex set of orbital occupations should be experimentally forbidden. In contrast to the classical computations of Borland and Dennis, we are not representing the quantum states by matrices but rather preparing quantum states directly, which allows us to measure the orbital occupations \textcolor{black}{experimentally}.

\section*{Results}

\subsection*{Measurement of 1-RDM Eigenvalues}

Pauli observed in 1926 that for quantum systems of fermion particles such as electrons the occupation $n_{i}$ of each spin orbital must obey the following inequalities:
\begin{equation}
0 \le n_{i} \le 1 ,
\end{equation}
known as the Pauli exclusion principle or Pauli constraints. In 1963 Coleman proved mathematically that these constraints plus a normalization constraint in which the occupation numbers sum to $N$ are necessary and sufficient for the occupation numbers $n_{i}$ to represent at least one ensemble state of $N$ electrons\cite{Coleman1963}.  Borland and Dennis in 1972, however, discovered that there exist additional conditions on the occupation numbers for the representation of at least one pure state of $N$ electrons, which are presently known as the generalized Pauli constraints\cite{Borland1972}.  A pure state is a quantum state that is describable by a single wave function.  Borland and Dennis found the following generalized Pauli constraints for three electrons in six orbitals:
\begin{equation}\label{bdinequal}
n_{5} + n_{6} - n_{4} \ge 0
\end{equation}
where
\begin{equation}\label{bdequal1}
n_{1} + n_{6}  = 1
\end{equation}
\begin{equation}\label{bdequal2}
n_{2} + n_{5}  = 1
\end{equation}
\begin{equation}\label{bdequal3}
n_{3} + n_{4}  = 1 ,
\end{equation}
where $n_{i}$ are the  \textcolor{black}{natural-orbital} occupations ordered from largest to smallest.

To test the generalized Pauli constraints on a quantum computer, we prepare an initial pure state $| \Psi_{0}(123) \rangle$ of 3 fermions in 6 orbitals and perform arbitrary unitary transformations ${\hat U}_{i}$ of the initial state to generate a set of random pure states $| \Psi_{i}(123) \rangle$ of 3 fermions in 6 orbitals
\begin{equation}
| \Psi_{i}(123) \rangle = {\hat U}_{i} | \Psi_{0}(123) \rangle .
\end{equation}
We measure the matrix elements of the 1-RDM of each state $| \Psi_{i}(123) \rangle$ generated on the quantum computer and check each 1-RDM to verify satisfaction of the generalized Pauli constraints for 3 fermions in 6 orbitals.  The 1-RDM is diagonalized on a classical computer and the eigenvalues (natural occupation numbers) are inserted into the generalized Pauli constraints to check for satisfaction.  The eigenvalues of the 1-RDM, ordered from highest to lowest, form a special type of convex set with ``flat'' sides known as a polytope.  The boundaries or ``flat'' sides of the polytope are determined by the Pauli and generalized Pauli constraints.  \textcolor{black}{Suppose the experimental data satisfies the Pauli constraints but not the generalized Pauli constraints,} then the smallest 3 eigenvalues of each measured 1-RDM will describe the Pauli polytope which is shown as the combined yellow and blue regions in Fig.~1.  On the other hand,  \textcolor{black}{suppose the experimental data obeys not only the Pauli constraints but also the generalized Pauli constraints,} then the smallest 3 eigenvalues of each measured 1-RDM will only describe the smaller generalized Pauli polytope, pictured as just the yellow region in Fig.~1. Comparison of the measured scatter plot of 1-RDM eigenvalues with these two polytopes provides  \textcolor{black}{an experimental means of observing the generalized Pauli constraints and thereby verifying their validity in a quantum system.}

\begin{figure}\label{fig:polytopes}
\begin{center}
\includegraphics[scale=0.40]{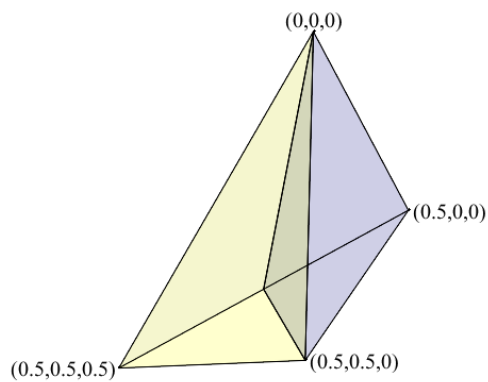}
\end{center}
\caption{{\bf Orbital occupations of the 1-electron reduced density matrix (1-RDM) form convex polytopes.} The eigenvalues (natural occupations) of the 1-RDM, ordered from largest to smallest, form a special convex set with ``flat'' sides known as a polytope.  Because the three smallest eigenvalues for a 3-electrons-in-6-orbitals state $(n_{4}, n_{5}, n_{6})$ determine the other eigenvalues, we can visualize the polytope in three dimensions.  The figure shows two polytopes: ({\em i}) the Pauli polytope of eigenvalues obeying the ordinary Pauli constraints (the combination of the yellow and blue regions) as well as ({\em ii}) the generalized Pauli polytope of eigenvalues obeying the generalized Pauli constraints (the yellow region only).  The plane separating the yellow and blue regions arises from the Borland-Dennis inequality shown in Eq.~(\ref{bdinequal}).}
\end{figure}

The system of 3 fermions in 6 orbitals can be expressed as a system of 3 qubits, allowing the above procedure to be simplified.  Because the occupations of each pair, $\{n_{1},n_{6}\}$,  $\{n_{2},n_{5}\}$, and  $\{n_{3},n_{4}\}$, must sum to one, the three electron systems has a one-to-one mapping to a system of three qubits.  The occupation numbers $n_{4}$, $n_{5}$, and $n_{6}$, which are eigenvalues of the 1-RDM, can be viewed as the eigenvalues $p_{1}$, $p_{2}$, and $p_{3}$ of the 1-qubit reduced density matrix of a three-qubits system.  Hence, for the three-qubits system the generalized Pauli constraints in Eqs.~\ref{bdinequal}-\ref{bdequal3} can be written as the single inequality
\begin{equation}\label{Higuchi}
p_{2}+p_{3}-p_{1} \ge 0 ,
\end{equation}
which was first obtained by Higuchi in 2002\cite{Higuchi2003}.  The Higuchi inequalities for $N$ qubits are a subset of the generalized Pauli constraints for $N$ electrons in $2N$ orbitals.  In the case of three electrons in six orbitals the Higuchi inequality is equivalent to the generalized Pauli constraint in Eq.~(\ref{bdinequal}) under the assumption that Eqs.~\ref{bdequal1}-\ref{bdequal3} hold.  In the quantum computation we exploit this relationship to represent the 3 electron in 6 orbital quantum system efficiently as a 3 qubit system with a compact fermionic mapping, described in the Methods section.  In this representation the violation of the Higuchi inequality in Eq.~(\ref{Higuchi}) is equivalent to the violation of the generalized Pauli inequality in Eq.~(\ref{bdinequal}).

The initial 3-qubit state is chosen to be the non-interacting state in which all 3 qubits are in their lower-energy (off) state.  The arbitrary unitary transformations are generated on the quantum computer from
\begin{equation}
| \Psi_{i}(123) \rangle =    C_1^3 R_y(\gamma) C_3^1 R_y(\beta) C_1^2 R_y(\alpha) | \Psi_{0}(123) \rangle ,
\end{equation}
where the parameters $\alpha$, $\beta$, and $\gamma$ in the Pauli rotation matrices $R_y(\alpha)$ are chosen randomly and the $C_i^j$ are controlled NOT (CNOT) gates. The rotation matrices are applied to the control qubit of the ensuing CNOT gate. It is known that any 3-qubit state can be prepared from the non-interacting state by a unitary transformation built from only 3 CNOT gates plus universal single-qubit gates\cite{Znidaric2008}. The above transformation generates states that span the most general entanglement class for the system and whose 1-qubit RDMs cover all possible real 1-qubit occupation numbers\cite{Acin2000,Sudbery2001,Walter2012}. Computations were also performed with a slightly more general transformation, discussed in the Methods and Supplementary Figure~1, albeit without a significant change in the results.  Because of the mapping between the 3-qubit and the 3-fermion-in-6-orbitals system, the measured occupations $p_{1}$, $p_{2}$, and $p_{3}$ are equivalent to the natural occupations (eigenvalues) $n_{4}$, $n_{5}$, and $n_{6}$ of the 1-RDM.  For the remainder of this work we primarily discuss the results in terms of the 3-fermion system.

\subsection*{Verification of Generalized Pauli Constraints}

The scatter plot of the measured 1-RDM natural occupations is shown in Fig.~2 relative to the Pauli polytope, the combination of the yellow and blue regions, and the smaller generalized Pauli polytope, only the yellow region.  Results show that the physical system only accesses the generalized Pauli polytope (yellow), defined by the generalized Pauli constraints.  None of the natural occupation numbers lie in the part (blue) of the Pauli polytope which is forbidden by the Borland-Dennis constraint.  Therefore, the experimental data from the quantum computer verifies the generalization of the Pauli exclusion principle for pure states.  Supplementary Figure~2  visualizes the effect of error on the occupation numbers.  We observe that the errors consistently push the triplet of occupations into the generalized Pauli polytope, which reflects upon the nature of the quantum noise. Despite the presence of quantum noise the fundamental result is statistically robust\cite{Walter2012}. The Pauli polytope is twice the size of the generalized Pauli polytope.  Consequently, without further restrictions beyond the ordinary Pauli constraints the probability of a randomly prepared state being in the yellow region would be one out of two.  The probability of $n$ random states being on the yellow side, therefore, would be $1/2^{n}$.  With $n$ being approximately 60, we observe that the probability of measuring all 60 points within the yellow region would be $1/2^{60}$ or a highly improbable one in one quintillion.  Hence, we have verified the generalized Pauli exclusion principle by quantum computer \textcolor{black}{to a high degree of confidence}.

\begin{figure}\label{figure2}\begin{center}
 \includegraphics[scale=0.40]{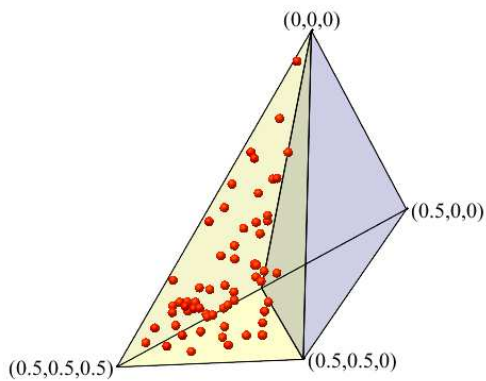}\end{center}
 \caption{{\bf Measured orbital occupations verify generalized Pauli principle.} The scatter plot of the three lowest measured 1-electron reduced density matrix (1-RDM) eigenvalues $(n_{4}, n_{5}, n_{6})$ is shown relative to the Pauli polytope, the combination of the yellow and blue regions, and the smaller generalized Pauli polytope, only the yellow region.  Results show that none of the eigenvalues lie in the part (blue) of the Pauli polytope which is forbidden for pure quantum states by the Borland-Dennis constraint.}
\end{figure}

Despite the restrictions on the pure-state observables from the generalized Pauli constraints, the set of realizable 1-RDMs exhibits all limits of quantum behavior including the mean-field limit as well as the strong-electron-correlation limit including phenomena like superconductivity.  Figure 3 shows examples of chemical systems which widely vary in the degree of correlation present. As observed in previous work, for 3-electron-in-6-orbital systems the ground-state triplet of occupation numbers $(n_{4}, n_{5}, n_{6})$ lies on the Borland-Dennis generalized Pauli constraint, separating pure and ensemble states in the set of 1-RDMs.  Excited-state natural occupation numbers of these systems, in contrast, can lie on the generalized Pauli constraint like the ground states or elsewhere in the polytope, reflecting substantial variations in the one-electron properties of excited states\cite{Chakraborty2014}.   \textcolor{black}{The 3 lowest occupation numbers $(n_{4}, n_{5}, n_{6})$ can be readily expressed in the natural-orbital basis set in terms of single, double, and triple excitations from the single determinant $| \phi_{1} \phi_{2} \phi_{3} \rangle$, composed of the 3 most occupied natural orbitals; for $n_{u}$ where $u$ indicates the index of one of the unoccupied orbitals of the reference determinant we have
\begin{equation}
n_{u} = \sum_{i}{|c_{i}^{u}|^{2} } + \sum_{ija}{|c_{ij}^{au}|^{2} } + \sum_{ijkab}{|c_{ijk}^{abu}|^{2}}
\end{equation}
in which $i$, $j$, and $k$ denote indices of natural orbitals in the reference determinant and $a$ and $b$ denote indices of natural orbitals not in the reference determinant and $c_{i}^{u}$, $c_{ij}^{au}$, $c_{ijk}^{abu}$ are the single-, double-, and triple-excitation coefficients.   From the formula we observe that when $(n_{4}, n_{5}, n_{6})$ equals $(0,0,0)$, all of the excitation coefficients vanish, and the quantum state is the reference determinant $| \phi_{1} \phi_{2} \phi_{3} \rangle$.  All other points in the polytope have contributions from one or more of the excitation coefficients.  The triple excitations vanish when the natural occupation numbers $(n_{4}, n_{5}, n_{6})$ are pinned to the Borland-Dennis inequality\cite{Klyachko2006,Altunbulak2008,SB2016}.}  The triplet of occupation numbers from ground and excited states tends to gravitate towards the boundaries of the polytope, reflecting the interplay between the energy optimization and the restrictions imposed on fermions from the Pauli constraints.

\begin{figure}\label{figure3}
\begin{center}
\includegraphics[scale=0.40]{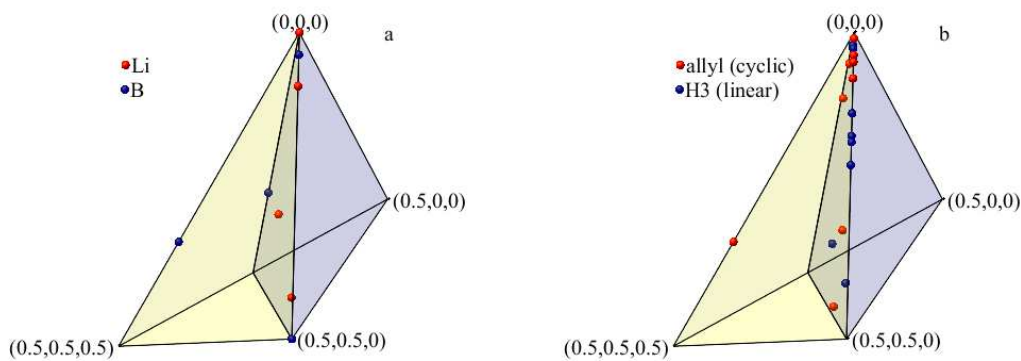}
\end{center}
\caption{{\bf Orbital occupations of correlated 3-electron atoms and molecules.} The lowest three eigenvalues (natural occupations) of the 1-electron reduced density matrix (1-RDM) for ground and excited states of 3-electron-in-6-orbital atoms and molecules $(n_{4}, n_{5}, n_{6})$ are shown: {\bf a} Lithium atom (red) and the $\pi$ orbitals of the Boron atom (blue). {\bf b} Allyl radical in the cyclic form (C$_3$H$_3$, red) and linear H$_3$ (blue).  The triplet of occupation numbers from ground and excited states tends to gravitate towards the boundaries of the polytope, reflecting the interplay between the energy optimization and the restrictions imposed on fermions from the Pauli constraints.  These eigenvalues were calculated by full configuration interaction on a classical computer.}
\end{figure}

\section*{Discussions}

The generalized Pauli constraints, discovered by Borland and Dennis on a classical computer, have been verified through calculations with a quantum computer.  While a quantum state is represented by vectors and matrices on a classical computer, it is actually prepared and manipulated on a quantum computer.  Consequently, the  presented computations can be interpreted as experiments that are measuring the generalized Pauli constraints' satisfaction by the correlated quantum states generated on the computer.  Using random unitary transformations to generate a statistical sample of the possible states, we observe that all of the states satisfy the generalized Pauli constraint, \textcolor{black}{experimentally verifying} the validity of the constraint \textcolor{black}{to a high degree of confidence}.  The generalized Pauli constraints have profound consequences for quantum phenomena.  First, their existence proves fundamental limitations on quantum entanglement's ability to emulate one-electron occupations and properties of ensemble quantum states.  Second, they imply the Higuchi inequalities and thereby place significant restrictions on many-qubit systems in pure states.  These restrictions may provide an exploitable resource for efficient error correction.  Finally, their recent extension to the two-electron reduced density matrix (2-RDM) provides new conditions for the 2-RDM to represent at least one many-electron pure-state quantum system\cite{Mazziotti2016}.  Such conditions can potentially be used to enhance the direct variational calculation of the 2-RDM\cite{Mazziotti2016b}, which is applicable to treating strongly correlated atomic and molecular quantum systems at a polynomial-scaling computational cost\cite{MM2018,SM2018,SM2018b}.

\section*{Methods}

We include details on the quantum algorithm used in the article and its variants, the method for generating the parameters, the quantum tomography of the one-electron reduced density matrix, and relevant details on the experimental quantum device used.

\subsection*{Quantum Algorithms} 

Two algorithms were used in this work. Both utilize 3 CNOT gates, which allow for the whole set of \textcolor{black}{3-}qubit occupation numbers to be spanned with suitable single qubit rotations. A single CNOT \textcolor{black}{can form a biseparable} system, while two CNOT gates can reach the GHZ state and spread a far range of the polytope, but three CNOT gates are required to \textcolor{black}{
saturate the Higuchi inequality}\cite{Higuchi2003,Znidaric2008,Acin2000,Sudbery2001}. Complex \textcolor{black}{unitary transformations}
 can be used, \textcolor{black}{increasing}
 the set of all possible quantum states, but \textcolor{black}{these transformations span} the same set of occupations as real rotations.


The first algorithm takes a minimalistic approach, and spans the polytope by parameterizing only three rotations. The algorithm (applied right to left) is as follows:
\begin{align}\label{alg1}
\mathbb{U}_a^1 = &  C_3^2 R_{y,3}(\theta_3)  C_1^2 R_{y,1}(\theta_2)  C_1^3 R_{y,1}(\theta_1),
\end{align}
\textcolor{black}{where}
$R_{y,i}$ refers to a qubit rotation around the y-axis of the Bloch Sphere onto the i$^{th}$ qubit (later we drop the $y$, as we only use $R_y$ rotations)\textcolor{black}{, defined as:}
\begin{align}
R_{y,i}(\theta) = \begin{pmatrix} \cos(\frac{\theta}{2}) & -\sin(\frac{\theta}{2}) \\ \sin(\frac{\theta}{2}) & \cos(\frac{\theta}{2})
\end{pmatrix}.
\end{align} $C_i^j$ is \textcolor{black}{the standard}
 CNOT gate with $i$ control and $j$ target qubits. The control qubit is rotated prior to a CNOT transformation. The \textcolor{black}{sequence of transformations produce}
 a wave function of the form:
\begin{equation}
|\Psi_f \rangle = \alpha |000\rangle + \beta |011 \rangle + \gamma |101 \rangle + \delta |110 \rangle,
\end{equation}
where $\alpha$, $\beta$, $\gamma$, and $\delta$ are all functions of $\theta_1$, $\theta_2$, and $\theta_3$, and the wave function has no diagonal elements in the 1-RDM.


The second algorithm provides a over-redundant representation of the system, \textcolor{black}{
including some local degrees of freedom\cite{Sudbery2001}}, and involves rotations on both the target and control qubit prior to the rotation, and is as follows:

\begin{align}\label{alg2}
\mathbb{U}_a^2 = C_3^1 R_3(\theta_6)R_1(\theta_5) C_2^3 R_2(\theta_4)R_3(\theta_3) C_2^1 R_2(\theta_2)R_1(\theta_1),
\end{align}
Note that the CNOT identity
\begin{align}
C_1^2 = H_1 H_2 C_2^1 H_1 H_2
\end{align}
implies that by performing \textcolor{black}{certain} single-qubit rotations on both qubits, we somewhat eliminate the dependence of the qubit orderings with a generic rotation. Of course the wave function is of a more general form, involving all 8 qubit states, and so the off-diagonal elements need to be determined.

\subsection*{Generating Parameters}

The parameters for each run were obtained by classically \textcolor{black}{simulating} the circuit over a range of parameters (from 0$^\circ$ to 45$^\circ$ in 0.1$^\circ$ intervals), calculating the theoretical eigenvalues after applying a given algorithm, and then either: 1) adding the point to a set if a point was further from all other points by a set distance, or; 2) discarding the point. Thus for the first algorithm we sampled 90 million points, and for the second, 8 quadrillion. The minimum distance for inclusion was 0.075. Only 62 and 59 points were selected for the first and second respectively. More points could have been used, but \textcolor{black}{this would not have provided additional clarity, due to the scale of the shifts from errors}. In general, unbiased random sampling did not yield `graphically' uniform sets, as there is a tendency to cluster in certain areas of the polytope, and would lead to lots of runs being only in a particular region. Thus, the above method was used.

\subsection*{Quantum Tomography of the 1-RDM}

The quantum tomography of the 3-electrons-in-6-orbitals system is simplified by its mapping to a system of 3 qubits.  We derive the unitary transformations required to perform quantum tomography of the 1-RDM for a completely general many-electron state and outline the simplified application of this tomography for the 3-electrons-in-6-orbitals system.


Consider the unitary transformation in terms of the rotation angle $\phi$
\begin{equation}
\label{eq:U1}
U = e^{\phi\hat{\alpha}}
\end{equation}
where
\begin{equation}
\hat{\alpha} = (\hat{a}^\dagger_i \hat{a}^{}_j - \hat{a}^\dagger_j\hat{a}^{}_i)
\end{equation}
where ${\hat a}^{\dagger}_{i}$ and ${\hat a}_{i}$ are second-quantized operators that create and annihilate an electron in orbital $i$.  We can express this unitary transformation $U$ in the following closed form:
\begin{equation}
e^{\phi\hat{\alpha}} = \hat{\beta}\cos\phi + \hat{\alpha}\sin\phi  + (1-\hat{\beta}),
\end{equation}
where
\begin{equation}
\hat{\beta} = {\hat a}_{i}^{\dagger} {\hat a}_{i} + {\hat a}_{j}^{\dagger} {\hat a}_{j} + {\hat a}_{i}^{\dagger} {\hat a}_{j}^{\dagger} {\hat a}_{i} {\hat a}_{j} + {\hat a}_{j}^{\dagger} {\hat a}_{i}^{\dagger} {\hat a}_{j} {\hat a}_{i}.
\end{equation}
Using the Baker-Campbell-Hausdorff expansion, we evaluate the unitary transformation of the projection operator ${\hat M}$ for measuring the occupation of the $i^{\rm th}$ orbital
\begin{align}
&e^{-\phi\hat{\alpha}} \hat{M} e^{\phi\hat{\alpha}} = i^\dagger i + \frac{\hat{\zeta}}{2}\sin2\phi - \hat{\eta} \sin^2 \phi
\end{align}
where
\begin{equation}
{\hat M} = {\hat a}^{\dagger}_{i} {\hat a}_{i}
\end{equation}
and
\begin{eqnarray}
\hat{\zeta} & = & {\hat a}_{i}^{\dagger} {\hat a}_{j} + {\hat a}_{j}^{\dagger} {\hat a}_{i} \\
\hat{\eta} & = & {\hat a}_{i}^{\dagger} {\hat a}_{i} - {\hat a}_{j}^{\dagger} {\hat a}_{j} .
\end{eqnarray}
 If we set the angle $\phi$ to $\pi / 4$ and take the expectation value with respect to the quantum state $| \Psi \rangle$, we obtain
\begin{equation}
\langle \Psi | e^{-\frac{\pi}{4}\hat{\alpha}}\hat{M}e^{\frac{\pi}{4}\hat{\alpha}} | \Psi \rangle = \frac{1}{2}\left (^1D^i_i+^1D^j_j+^1D^i_j+^1D^j_i \right)
\end{equation}
in which $^{1} D^{i}_{j}$ denotes an element of the 1-RDM.  Therefore, by taking expectation values for the diagonal elements of the family of 1-RDMs from these unitary transformations, we obtain information about all of the elements of the 1-RDM from which we can obtain its off-diagonal elements.  While these transformations are sufficient to determine the real part of the 1-RDM, in the case that the 1-RDM has a non-vanishing imaginary component, we must also consider the unitary transformations where ${\hat \alpha}$ is defined as
\begin{equation}
\hat{\alpha} = i (\hat{a}^\dagger_i \hat{a}^{}_j + \hat{a}^\dagger_j\hat{a}^{}_i).
\end{equation}
By an analogous procedure we obtain
\begin{equation}
\langle \Psi | e^{-\frac{\pi}{4}\hat{\alpha}}\hat{M}e^{\frac{\pi}{4}\hat{\alpha}} | \Psi \rangle = \frac{1}{2}\left (^1D^i_i+^1D^j_j+i (^1D^i_j-^1D^j_i) \right).
\end{equation}
from which we can extract the imaginary off-diagonal elements of the 1-RDM.


Because of Eqs.~(\ref{bdequal1}-\ref{bdequal3}) of the Borland-Dennis constraints the 6 natural orbitals of the 3-electrons-in-6-orbitals system can be paired to form 3 qubits where each qubit is a two-level system sharing an electron.  We can restrict the one-body unitary transformation in Eq.~(\ref{eq:U1}) to indices $i$ and $j$ representing orbitals of the same qubit because other choices violate the restriction of one electron per qubit.  Hence, the one-body fermionic unitary transformation can be represented as a unitary transformation of a single qubit
\begin{equation}
U =
\begin{pmatrix}
\cos \phi & -\Gamma \sin \phi \\ \Gamma\sin \phi & \cos \phi ,
\end{pmatrix}
\end{equation}
where $\Gamma$ is a global phase from the antisymmetry of electrons.  This mapping to qubits can be viewed as a member of the family of compact mappings~\cite{Aspuru-Guzik2005}.  Practically, this transformation is implemented in this study for $\phi = \pi/4$ as a product of the Hadamard gate $H$ and a Pauli-Z gate $Z$.  If $\Gamma=1$ and $\phi = \pi/4$, then $U =HZ$, and if $\Gamma=1$ and $\phi = \pi/4$, then $U =ZH$.

\subsection*{Quantum Computation}

In this work we used the IBM Quantum Experience devices (ibmqx4 and ibmqx2), available online, \textcolor{black}{in particular the} 5-transmon quantum computing device\cite{Koch2007}. These cloud accessible quantum devices are fixed-frequency transmon qubits with co-planer waveguide resonators\cite{Koch2007,Chow2011}. Experimental calibration for these devices is included below, and connectivity is specified there. Additionally, results are included in Supplementary Tables~1--8.

For our work, we tested varying number of measurements, and found that no significant decrease in the error of a run occurred by using more than 2048 measurements, and so the primary and secondary experiments spanning the polytope utilized 2048 measurements of the quantum algorithm. Additionally, for showing how error shifts the occupation of a pure state, we utilized 1024 measurements, as in Supplementary Figure~2.

To show this, let the target variable be the distance of the measured occupation to the \textcolor{black}{ideal} point. Let the mean of a run with 8192 measurements be the population mean, then define a random variable $\chi$ representing the distance from the ideal point after 32 measurements (2$^2$ times the number of possible qubit states). For a point lying within the GPC polytope, generated with equation \ref{alg1}, where $\theta_1=43.0^\circ$, $\theta_2=3.0^\circ$, and $\theta_3=39.0^\circ$, we find $\mu_\chi = 0.059$ and $\sigma_\chi = 0.056$. For $2048$ and $1024$ iterations, this yields a standard error of 0.007, and 0.010, respectively, and the 95\% confidence interval is then $\pm 0.014$, and $\pm 0.020$, respectively. For other points similar values were seen, typically with higher average $\mu_\chi$. Considering the scale of our problem, where shifts from the ideal occupation are closer to 0.1, and direction plays a significant role, one would not expect a significant difference in results upon increased iterations. The three qubits with the lowest two-qubit gate error and the correct ordering of the qubit algorithm was used for the device connectivity.

\subsection*{Quantum Computer Calibration}

Calibration data from the primary and secondary algorithms as provided by IBM's quantum computer is presented in Table~1~\cite{Sparrow2018}. \\

\begin{table}

\caption{{\bf Calibration Data}}

\begin{tabular}{ccccccc} \hline \hline
Device: & ibmqx2 (``Sparrow'') \\
Calibration Date: & 2-23-2018 \\
Temperature ($^\circ$ K) : & 0.0164 \\
Version: & 3.0 \\
Buffer (ns): & 6.7 \\
Gate time (ns): & 83.3 \\
\hline
Qubit: & 0 & 1 & 2 & 3 & 4 \\
\hline
T2 ($\mu$s) : 	&	41.5 	& 55.3		& 67.1	&	69.8 & 44.2 \\
$f$ (GHz): 		&	5.27603 & 5.21224	& 5.01541&	5.28059 & 5.07117\\
T1 ($\mu$s): 	&	59.4 	& 67.8		& 68.9	& 	48.9 & 66.0 \\
Gate Error ($10^{-3}$): 	&	1.98 & 1.29	& 1.98 & 1.63 & 0.94 \\
Readout Error ($10^{-3}$): 	&	45 	& 36		& 20	 &	16 & 25 \\
\hline
Multi-Qubit: & 01 & 02 & 12 & 32 & 34 & 42 \\
\hline
Error ($10^{-3}$) & 34.6 & 40.7 & 32.6 & 27.6 & 22.3 & 26.6 \\
 \hline \hline
\end{tabular}

\end{table}

\subsection*{Electronic Structure Calculations}

Molecular geometries for C$_3$H$_3$ were taken from the Computational Chemistry Comparison and Benchmark Database\cite{NIST}, and  \textcolor{black}{for H$_3$ were calculated with Gaussian 09 with the coupled cluster singles and doubles method (CCSD)\cite{Frisch2010}}. The basis set used Slater-type orbitals with three Gaussians (STO-3G). Electron integrals were obtained from General Atomic and Molecular Electronic Structure System (GAMESS)\cite{Schmidt1993}. Maple~\cite{Maple} was used to perform a full configuration interaction (FCI) calculation with a QR method\cite{Golub} for ground and excited states including their 1-RDMs.

\section*{Data Availability}

\noindent The data that support the finding of this study are presented in Supplementary Tables 1--8. Any additional data are available from the corresponding author upon reasonable request.

\section*{Acknowledgments}

\noindent D.A.M. gratefully acknowledges the Department of Energy Grant, the U. S. National Science Foundation Grant CHE-1565638, and the U.S. Army Research Office (ARO) Grant W911NF-16-1-0152.

\section*{Author contributions}

\noindent D. A. M. conceived of the research project.   S. E. S. 	and D. A. M. developed the theory.  D. I. S. provided guidance on performing the computations on a quantum computer.  S. E. S. performed the calculations.  S. E. S., D. I. S., and D. A. M. discussed the data and wrote the manuscript.

\section*{Competing interests}

\noindent The authors declare no competing interests.


\end{document}